\documentclass[aps,prb,showpacs,twocolumn,amssymb]{revtex4}

\usepackage{amsmath}
\usepackage{subfigure}
\usepackage{graphicx}

\input epsf

\def\prb{Phys. Rev. B }
\def\prl{Phys. Rev. Lett. }

\def\be{\begin{equation}}
\def\ee{\end{equation}}
\def\ba{\begin{eqnarray}}
\def\ea{\end{eqnarray}}
\def\ie{{\it i.e.} }

\def\etal{{\it et al.} }

\def\124{YBa$_2$Cu$_4$O$_8$ }

\def\C60{A$_x$C$_{60}$ }

\begin{document}

\title{Superfluid stiffness renormalization and critical temperature enhancement in a composite superconductor}

\author{Gideon~Wachtel, Assaf~Bar-Yaacov and Dror~Orgad}

\affiliation{Racah Institute of Physics, The Hebrew University, Jerusalem 91904, Israel}

\date{\today}

\begin{abstract}

We study a model of a composite system constructed from a "pairing layer" of disconnected attractive-$U$
Hubbard sites that is coupled by single-particle tunneling, $t_\perp$, to a disordered metallic layer.
For small inter-layer tunneling the system is described by an effective long-range $XY$ phase model whose
critical temperature, $T_c$, is essentially insensitive to the disorder and is exponentially suppressed by
quantum fluctuations.
$T_c$ reaches a maximum for intermediate values of $t_\perp$, which we calculate using a combination of
mean-field, classical and quantum Monte Carlo methods. The maximal $T_c$ scales as a fraction of the
zero temperature gap of the attractive sites when $U$ is smaller than the metallic
bandwidth, and is bounded by the maximal $T_c$ of the two-dimensional attractive Hubbard model for
large $U$. Our results indicate that a thin, rather than a thick, metallic coating is better suited
for the enhancement of $T_c$ at the surface of a phase fluctuating superconductor.

\end{abstract}


\maketitle

\section{Introduction}
\label{intro}

Recently, composite systems made of a metal overlaying a superconductor with low phase stiffness
have received theoretical attention\cite{Ours,Okamoto,Ehud} after experiments have demonstrated
that the superconducting transition temperature, $T_c$,
can be enhanced in bilayers consisting of underdoped and overdoped cuprates.\cite{Yuli,Gozar}
In particular, Ref. \onlinecite{Ours} considered a model of a pairing layer with zero phase stiffness,
constructed from disconnected attractive-$U$ Hubbard sites, that is coupled via single-particle
tunneling $t_\perp$ to a metallic noninteracting layer.
As $t_\perp$ increases from zero, phase coupling between the pairing sites is established
by Josephson tunneling through the metallic layer. At the same time, however,
the pairing strength is diminished owing to the proximity effect induced by the same
delocalization events. Consequently, $T_c$ reaches a maximum at an intermediate value of
$t_\perp$, where both these trends attain a simultaneous optimum. It was shown, within a
mean-field approximation, that the maximal $T_c$ approaches the limit set by the
pairing scale of the Hubbard sites when the metallic bandwidth, $8t$, is much larger than $U$.
Here we wish to reexamine these results using more accurate analytical and numerical techniques.

An additional goal of the present work is to study the effects of disorder in the metallic
layer on $T_c$. Owing to Anderson\cite{anderson}, it is well known that $T_c$ of a dirty BCS
$s$-wave superconductor is essentially insensitive to disorder as long as $T_c$ is much larger
than the local mean level spacing. However, BCS theory ignores fluctuations in the superconducting
order parameter and one expects deviations from Anderson's result when fluctuations are large.
Indeed, phase fluctuations dominate the physics in strongly disordered and granular
systems,\cite{Ma,Fisher,chak,Ghosal98,Beloborodov,Dubi,borissteve}
where they can reduce $T_c$ to zero. In the bilayer that we study the disorder resides away
from the pairing sites and is therefore expected to have only mild consequences for pairing.
However, since phase stiffness is established by Josephson tunneling through the metal
it is interesting to study the manner in which it is affected by the random potential.

Our results indicate that although the general picture obtained in Ref. \onlinecite{Ours}
holds, it requires some modifications. First, the analysis of Ref. \onlinecite{Ours} relied on
a Bogoliubov de-Gennes (BdG) mean-field treatment to calculate the phase stiffness.
While this approximation takes into account the suppression of the phase stiffness by
thermally excited quasiparticles, it ignores the renormalization of the stiffness by thermal and quantum phase
fluctuations. The latter have little effect in the nearest-neighbor $XY$ model and the two-dimensional
attractive Hubbard model\cite{Paiva04}, where using the unrenormailzed stiffness reproduces the correct
$T_c$ to within 40\%. We show that this is not the case in the presence of long range phase couplings.
Such couplings, which extend up to the thermal length of the metal, occur in the small $t_\perp$ regime
of the bilayer. Under these circumstances classical phase fluctuations on scales
smaller than the thermal length lead to a rapid decrease of the stiffness and therefore
to a much lower $T_c$ than is anticipated from the unrenormalized stiffness. Even more important
are the quantum phase fluctuations in the small $t_\perp$ regime, which lead to an exponential
suppression of $T_c$ relative to its mean-field value.

Secondly, using quantum Monte Carlo and classical Monte Carlo mean-field techniques we find that the
highest $T_c$ (maximized over $t_\perp$), is smaller by a factor 3-4 than the mean-field prediction
for small and intermediate $U/t$. For large $U/t$ it is bounded by the maximal $T_c$ of the two-dimensional
attractive Hubbard model. Our calculations reveal that the maximum is largely governed by classical
phase fluctuations.

For small inter-layer tunneling the primary effect of the disorder is to decrease the range over which
coherent phase coupling is mediated through the metal. However, as long as the metal remains in the diffusive
regime, this has a negligible effect on $T_c$. Using a mean-field approach to estimate the effects of disorder
away from the small $t_\perp$ regime yields a weak correction to the maximal $T_c$ at small disorder
strength. Nevertheless, once the disorder becomes large in comparison to the hopping amplitude
in the metallic layer the maximal $T_c$ is significantly suppressed. We conclude the paper
with a short discussion of the relevance of these insights to attempts to enhance $T_c$ at the
surface of a phase fluctuating superconductor.

\section{The model and its analysis in the small $t_\perp$ limit}

\subsection{The model}

\begin{figure}[t]
\includegraphics[angle=0,width=\linewidth,clip=true]{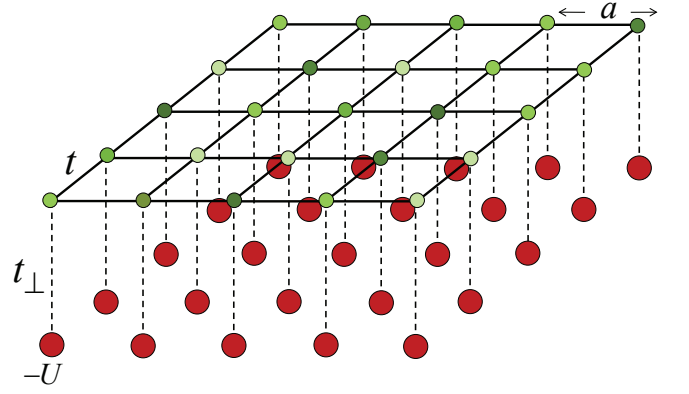}
\caption{The model: A square lattice of disconnected attractive-$U$ Hubbard sites is
coupled via single-particle tunneling $t_\perp$ to a noninteracting
tight-binding layer with on-site random disorder.}
\label{model-fig}
\end{figure}

We consider a bilayer, see Fig. \ref{model-fig}, consisting of a noninteracting disordered upper layer
and a lower layer of disconnected negative-$U$ Hubbard sites. Each layer contains $N=L^2/a^2$ sites in
a square array with lattice constant $a$.
Neighboring sites on the two layers are connected via single particle tunneling, $t_\perp$. We denote
by $c_{i\sigma}$ and $f_{i\sigma}$ the annihilation operator of an electron with spin $\sigma$ on the
$i$th site of the upper and lower layer, respectively.
The imaginary time action describing the model is
\ba
\label{model}
\nonumber
\!\!\!\!\!\!S\!\!\!\!&&=\int_{0}^{\beta}d\tau\left[ \sum_{i,\sigma}c_{i\sigma}^{\dagger}\left(
\partial_{\tau}-\mu+\epsilon+V_i \right)c_{i\sigma}\right. \\
\nonumber
\!\!\!\!\!\!&&-\; t\sum_{\left\langle i,j\right\rangle ,\sigma}
\left(c_{i\sigma}^{\dagger}c_{j\sigma}+{\rm H.c.}\right)
+\sum_{i,\sigma}f_{i\sigma}^{\dagger}\left(\partial_{\tau}-\mu\right)f_{i\sigma} \\
\!\!\!\!\!\!&&-\left. U\sum_{i}f_{i\uparrow}^{\dagger}
f_{i\uparrow}f_{i\downarrow}^{\dagger}f_{i\downarrow}-t_{\perp}\sum_{i,\sigma}
\left(c_{i\sigma}^{\dagger}f_{i\sigma}+{\rm H.c.}\right)\right].
\ea
Here $\beta=1/T$ is the inverse temperature, $\mu$ is the chemical potential and $t$ is the hopping amplitude
between sites in the upper layer, which contains a Gaussian random potential $V$ and a constant
potential $\epsilon$ used to adjust the Fermi energy away from any van-Hove singularities.

\subsection{Phase-only action}
\label{phase-action}

We proceed by decoupling the interaction term using a Hubbard-Stratonovich transformation
$-Uf_{i\uparrow}^{\dagger}f_{i\downarrow}^{\dagger}f_{i\downarrow}f_{i\uparrow}\rightarrow
\Delta^*_i\Delta_i/U+\Delta_i^* f_{i\downarrow}f_{i\uparrow}+
\Delta_i f^\dagger_{i\uparrow}f^\dagger_{i\downarrow}$, where $\Delta_i=|\Delta_i|e^{i\theta_i}$.
Gauge transforming $f_{i\sigma}\rightarrow f_{i\sigma} e^{i\theta_i/2}$, integrating out
the noninteracting layer and denoting $\Psi_i^\dagger=(f_{i\uparrow}^\dagger,f_{i\downarrow})$ leads to
\ba
\label{S1}
\nonumber
S&=&\int_0^\beta d\tau d\tau'\sum_{ij} \Psi_i^\dagger(\tau)[M^{(0)}+V^{(0)}+V^{(1)}]\Psi_j(\tau') \\
&+&\int_0^\beta d\tau \sum_i |\Delta_i(\tau)|^2, \\
\nonumber
\ea
with
\ba
\label{MVreal}
\nonumber
M^{(0)}&=&[\partial_\tau{\bf I}-\mu\mbox{\boldmath $\sigma$}_3+|\Delta_i(\tau)|\mbox{\boldmath $\sigma$}_1]\delta_{ij}\delta(\tau-\tau'),\\
\nonumber
V^{(0)}&=&\frac{i}{2}\partial_\tau\theta_i(\tau)\mbox{\boldmath $\sigma$}_3\delta_{ij}\delta(\tau-\tau'),\\
V^{(1)}&=&t_\perp^2e^{-i[\theta_i(\tau)-\theta_j(\tau')]/2}{\cal G}_{ij}(\tau-\tau')\mbox{\boldmath $\sigma$}_3,
\ea
where ${\cal G}_{ij}(\tau)$ is the Green's function
of the noninteracting layer, and $\mbox{\boldmath $\sigma$}$ are the Pauli matrices.

Next, we integrate out the degrees of freedom of the pairing layer. We do so perturbatively to second
order in $V^{(0)}$ and $V^{(1)}$. The time independent amplitude
$|\Delta_i|\equiv|\Delta_i|(\Omega_n=0)$ is essentially set by the zeroth order contribution
$(\beta/U)\sum_i|\Delta_i|^2-{\rm Tr}\ln M^{(0)}(|\Delta_i|)$ to the effective action,
which provides within a saddle point approximation the BCS gap equation for the decoupled Hubbard sites.
We assume that the pairing layer is close to half filling such that $\mu\ll|\Delta|$, and obtain as a result
the $T=0$ solution $|\Delta|\equiv\Delta_0\approx U/2$.
The ${\cal O}(t_\perp^2)$ contribution modifies the gap
equation and leads to a reduction of order $t_\perp^2/t$ of $\Delta_0$, reflecting
the proximity effect. In the following we assume that $t_\perp^2/t,T\ll\Delta_0$
and therefore neglect both this effect and any thermal or quantum fluctuations of the gap amplitude.

The action governing the phase fluctuations is derived in the appendix. There we show that its dominant
terms are
\ba
\label{Stheta}
\nonumber
\!\!\!\!\!\!\!S_\theta&=&\frac{1}{8\Delta_0} \int_0^\beta d\tau \sum_i (\partial_\tau\theta_i)^2 \\
\!\!\!\!\!\!\!&-& \int_0^\beta d\tau d\tau' \sum_{i,j} K(r_{ij},\tau-\tau')\cos[\theta_i(\tau)-\theta_j(\tau')].
\ea
The Kernel $K(r,\tau)$ decays exponentially for spatial separations $r_{ij}=|{\bf r}_i-{\bf r}_j|$ larger than the thermal length
$l_T$. This decay reflects the loss of coherence between the dynamical phases of the members of a pair that mediates
the phase coupling. Due to thermal smearing of the Fermi-Dirac distribution the difference between the energies of
the two electrons is of order $T$, which leads to loss of coherence after time $1/T$. In the clean limit where the
elastic mean free time of the metal satisfies $\tau_e^{-1}\ll T$ this coherence time translates to a distance
$l_T=v_F/T$ covered by the ballistically propagating electrons with Fermi velocity $v_F$. In the same clean
limit we find for $l_T\gg r\gg \xi=v_F/\Delta_0$, that the kernel is
\ba
\label{Klongclean}
K(l_T\gg r\gg \xi,\tau)=\frac{t_\perp^4 N_F a^4}{2\pi^2 v_F}\frac{1}{r}\frac{1}{(\Delta_0\tau)^2+(r/\xi)^2},
\ea
where $N_F$ is the density of states of the metallic layer at the Fermi energy. Up to logarithmic corrections the
behavior at shorter distances may be approximated by
\ba
\label{Kshortclean}
\nonumber
\!\!K(\xi\gg r\gg a,\tau)&=&\frac{t_\perp^4 N_F a^4}{2\pi^2 v_F}\frac{1}{r}\\
\!\!&\times&
\left[\frac{\pi^2}{4}e^{-2\Delta_0|\tau|}+\frac{(\Delta_0\tau)^2}{1+(\Delta_0\tau)^4}\right].
\ea

Owing to the reasons outlined above the phase coupling in the diffusive regime $\tau_e^{-1}\gg T$ exhibits a similar
exponential decay beyond the thermal length, which for a disordered metal with diffusion constant $D$ is given by
$l_T=\sqrt{D/T}$. In the important range $l_T\gg r\gg \xi=\sqrt{D/\Delta_0}$ the kernel behaves to within logarithmic
accuracy as
\ba
\label{Klongdiff}
K(l_T\gg r\gg \xi,\tau)=\frac{t_\perp^4 N_F a^4}{2\pi^2 D}\frac{1}{(\Delta_0\tau)^2+(r/2\xi)^4}.
\ea

\subsection{Classical phase fluctuations}
\label{thermal}

In the small $t_\perp$ limit the superconducting critical temperature, $T_c$, equals the phase ordering
temperature as determined by the action, Eq. (\ref{Stheta}). Being a finite temperature transition in a
two-dimensional system with finite range couplings it is clear that the phase ordering transition belongs
to the classical Berezinskii-Kosterlitz-Thouless (BKT) universality class. However, $T_c$ itself is
determined by both quantum (time-dependent) and classical (time-independent)
phase fluctuations. As we shall see, the quantum phase fluctuations can not be neglected and are in fact dominant for
small $t_\perp$, leading to exponential suppression of $T_c$. Albeit, we begin by considering the effects of the
classical fluctuations. We do so since the treatment of the classical fluctuations will reveal a lesson
concerning the renormalization of the phase stiffness which is generic to models with interactions that extend
well beyond nearest neighbors and since it will provide the tool to calculate the effects of the quantum fluctuations.

To this end, consider the case of a time-independent but space fluctuating $\theta$. Assuming $\beta\Delta_0\gg1$
and carrying out the time integration in Eq. (\ref{Stheta}) results in
\be
\label{F}
S_{\rm class}=-\beta\frac{t_\perp^4 N_F a^2}{\Delta_0^2}\sum_{i,j}f(r_{ij})\cos(\theta_i-\theta_j),
\ee
with phase couplings $f(r)$ given by
\ba
\label{f}
\begin{array}{ccc}
 & {\rm clean\;limit} & {\rm diffusive\;limit} \\
\xi\gg r\gg a,l_e & \frac{a^2}{4\xi r} & -\frac{1}{4}\left(\frac{a}{\xi}\right)^2
\ln\left(\frac{e^{\gamma-1/2}}{\sqrt{2}}\frac{r}{\xi}\right) \\ \\
\l_T\gg r\gg \xi & \frac{1}{2\pi}\left(\frac{a}{r}\right)^2 & \frac{1}{\pi}\left(\frac{a}{r}\right)^2 \\ \\
r\gg l_T & \frac{2a^2}{l_T r}e^{-2\pi r/l_T} & \left(\frac{a}{l_T}\right)^{\frac{3}{2}}\frac{(2\pi)^{1/4}}{\sqrt{r/a}}e^{-\sqrt{2\pi}\,r/l_T},
\end{array}
\ea
where $l_e=v_F\tau_e$ is the elastic mean free path and $\gamma$ is the Euler constant. In our region of interest $T\ll\Delta_0$ and
the physics is governed by the couplings in the range $l_T\gg r\gg \xi$. Thus, we are led to study the
following $XY$-type model
\be
\label{XYmodel}
H=-\frac{J}{2}\sum_{i,j}\left(\frac{a}{r_{ij}}\right)^2 e^{-r_{ij}/\lambda}
\cos(\theta_i-\theta_j),
\ee
with $J\sim t_\perp^4 N_Fa^2/\Delta_0^2$ and $\lambda\sim l_T$ as an effective description of the classical phase fluctuations.
Here, for sake of simplicity, we extended the $r^{-2}$ behavior of the coupling down to the short distance cutoff, ignoring the
crossover when $r<\xi$, see Eq. (\ref{f}). As we shall demonstrate, this leads to a negligible correction to $T_c$.

Since the phase ordering transition belongs to the BKT class its critical temperature is related to the universal
jump of the renormalized phase stiffness at criticality:
\be
\label{BKTcrit}
\rho_s(T_c)=\frac{2}{\pi}T_c.
\ee
In turn, the phase stiffness is calculated
from the free energy in the presence of a phase twist, $\phi$, per bond in the $x$ direction. That is, if
$H=-(1/2)\sum_{i,j}J_{ij}\cos[\theta_i-\theta_j+(x_i-x_j)\phi/a]$ then
\ba
\label{rhosform}
\nonumber
&&\!\!\!\!\!\!\!\!\!\!\!\!\!\!\!\!\rho_s(T)=\left.\left(\frac{a}{L}\right)^2\frac{\partial^2 F}{\partial\phi^2}
\right|_{\phi=0} \\
\nonumber
&&=\frac{1}{2L^2}{\Bigg \{}{\Bigg \langle}\sum_{i,j}(x_i-x_j)^2J_{ij}\cos(\theta_i-\theta_j){\Bigg \rangle} \\
&&-\frac{1}{2T}{\Bigg \langle}{\Bigg [}\sum_{i,j}(x_i-x_j)J_{ij}\sin(\theta_i-\theta_j){\Bigg ]}^2{\Bigg \rangle}{\Bigg \}},
\ea
where here $\langle\rangle$ denotes thermal averaging.

For the standard $XY$ model it is known that the BKT criterion (\ref{BKTcrit}) yields a fair estimate for $T_c$ even
when $\rho_s$ is replaced by the bare stiffness $\rho_s^0$, unrenormalized by vortices and longitudinal phase fluctuations.
Indeed, for that model $\rho_s^0=J$, as calculated from Eq. (\ref{rhosform}) using a uniform phase field $\theta_i=\theta$.
This gives the estimate $T_c=(\pi/2) J$, which is to be compared with the most recent numerical value\cite{Hasen05}
$T_c=0.8929 J$. One may, therefore, attempt to apply the same approximation to calculate the transition temperature
of model (\ref{XYmodel}). For this case one finds, using the fact $\lambda\gg a$, that
\be
\label{barerholong}
\rho_s^0=\frac{J}{2a^2}\int d^2r \left(\frac{x}{r}\right)^2 e^{-r/\lambda}=\frac{\pi}{2}
\left(\frac{\lambda}{a}\right)^2 J ,
\ee
and the estimate $T_c=(\pi\lambda/2a)^2J$.
Accordingly, the temperature dependence of $\lambda\sim l_T$ would imply $T_c\sim t_\perp^2/\Delta_0$ for the diffusive
system and $T_c\sim(t_\perp^4N_F v_F^2/\Delta_0^2)^{1/3}$ in the clean limit. The latter result was previously
derived in Ref. \onlinecite{Ours} within the same approximation.

\begin{figure}[t]
\includegraphics[angle=0,width=\linewidth,clip=true]{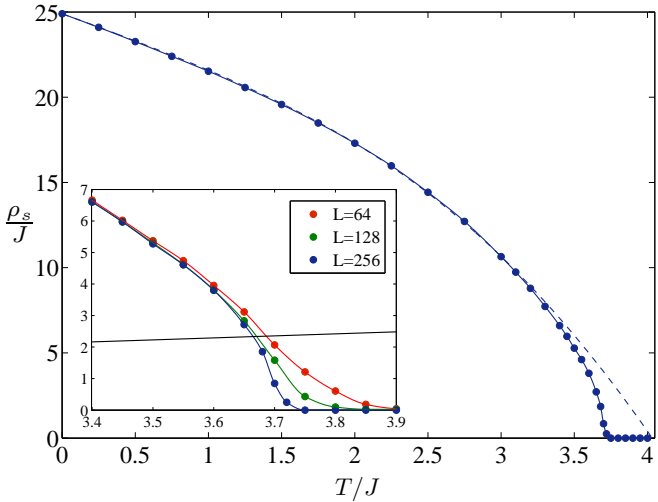}
\caption{The phase stiffness of the effective model, Eq.  (\ref{XYmodel}), with $\lambda/a=4$.
The dashed line is a fit (with $\lambda/a=3.897$, ${\bar J}/J=0.108$ and $z=8$) using the mean-field
approximation discussed in the text. The inset shows the size dependence of $\rho_s$ in the vicinity
of the transition, which occurs near the crossing of the stiffness curves with the solid black
line $\rho_s=(2/\pi)T$.}
\label{rhoT}
\end{figure}

\begin{figure}[t]
\includegraphics[angle=0,width=\linewidth,clip=true]{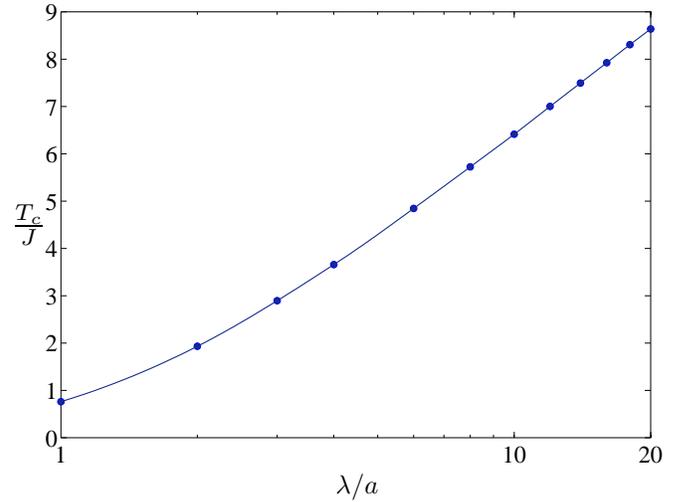}
\caption{The critical temperature of the effective model, Eq.  (\ref{XYmodel}), as function of
$\lambda/a$.}
\label{Tcplot}
\end{figure}

To test the validity of these estimates we calculated  $\rho_s(T)$ for the Hamiltonian, Eq. (\ref{XYmodel}),
via Monte-Carlo simulations of systems with up to $L=512$. For this purpose we used Eq. (\ref{rhosform}) and implemented
the Wolff algorithm\cite{Wolff}, which is easily generalized to include couplings beyond nearest-neighbor range.
Our results, see Fig. \ref{rhoT}, indicate that the phase stiffness develops a discontinuity, in accord with
the expected signature at a BKT transition. However, unlike the situation in the standard $XY$ model, $\rho_s(T_c)$
is massively renormalized down from its bare ($T=0$) value $\rho_s^0$. In fact, as shown by Fig. \ref{Tcplot},
this renormalization leads in our region of interest $\lambda/d\gg 1$ to $T_c\simeq 3.21J\ln(0.74\lambda/a)$ instead
of $T_c\sim J(\lambda/a)^2$. Therefore, using the above stated values of $J$ and $\lambda$ in terms of the parameters
of the original model we find the following estimate for $T_c$, based on thermal fluctuations alone
\be
\label{Tcclass}
T_{c,{\rm class}}\sim \frac{N_F a^2 t_\perp^4}{\Delta_0^2}\ln\left(\frac{\sqrt{N_F \eta}\Delta_0}
{N_Fa^2 t_\perp^2}\right)+{\cal O}\left(t_\perp^4\ln\ln t_\perp^{-1}\right),
\ee
where $\eta=D,v_F a$ in the diffusive and ballistic regimes, respectively.


In order to acquire an insight into these findings let us consider the following coarse-grained model where we divide
the system into blocks containing $(\lambda/a)^2$ unit spins. Each spin is assumed to interact with all the other spins
in its own block and in the $z$ neighboring blocks. To make analytical progress we replace the original coupling in
our model, which decays as $1/r^2$, by its average over a block. Hence, if we denote by $\mathbf{s}_i^I$,
$i=1\ldots (\lambda/a)^2$, the spins on the $I$th block, and by $\mathbf{m}_I=\sum_i \mathbf{s}_i^I$ the total
(super)spin on that block, we are led to study the model
\be
\label{effmodel}
{\bar H}=-\frac{\bar J}{2}\sum_I \mathbf{m}_I^2 - {\bar J}\sum_{\langle I,J\rangle} \mathbf{m}_I \cdot \mathbf{m}_J ,
\ee
where $\langle I,J\rangle$ denotes the blocks that couple to block $I$ and
\be
{\bar J}\approx J\left(\frac{a}{\lambda}\right)^2\int_a^\infty d^2r\, \frac{e^{-r/\lambda}}{r^2}
\approx J\left(\frac{a}{\lambda}\right)^2\ln\left(\frac{\lambda}{a}\right).
\ee

The coarse-grained model, Eq. (\ref{effmodel}), is a model of soft superspins whose average length
$M=\langle| \mathbf{m}|\rangle$ is determined by the intra-block phase fluctuations. As long as
fluctuations in $M$ are ignored the system can be viewed as an ordinary $XY$ model for unit superspins with
coupling ${\bar J}M^2$. Hence, we are interested in calculating the temperature dependence of the latter.
To this end, we continue to treat the intra-block fluctuations exactly but treat the inter-block coupling
in mean-field approximation using the following effective single block Hamiltonian
\be
\label{MFH}
{\bar H}_{MF}=-\frac{\bar J}{2}\mathbf{m}^2-{\bar J}z\mathbf{M} \cdot \mathbf{m},
\ee
with the self-consistency condition $\mathbf{M} =\langle\mathbf{m}\rangle_{{\bar H}_{MF}}$.

Using the Hubbard-Stratonovich transformation\cite{Antoni}
\be
\label{HST}
e^{\beta{\bar J}\mathbf{m}^2/2}=\frac{1}{\pi}\int_{-\infty}^{\infty}d^2 y e^{-\mathbf{y}^2
+\sqrt{2\beta{\bar J}}\mathbf{m}\cdot\mathbf{y}},
\ee
we are able to write the mean-field partition function as
\be
\label{ZMF1}
Z_{MF}=\int_{-\infty}^{\infty}\frac{d^2 y}{\pi}\prod_{i=1}
^{(\lambda/a)^2}\int_0^{2\pi}d\phi_i e^{-\mathbf{y}^2+|\mathbf{x}|\cos\phi_i},
\ee
where $\phi_i$ is the angle between $\mathbf{s}_i$ and
$\mathbf{x}=\sqrt{2\beta{\bar J}}\mathbf{y}+z\beta{\bar J}\mathbf{M}$. Carrying out the integrations
over the $\phi_i$s and the direction of $\mathbf{x}$ we arrive at
\ba
\label{ZMF2}
\nonumber
&&\!\!\!\!\!\!\!\!\!\!\!\!\!\!Z_{MF}=\frac{(2\pi)^{(\lambda/a)^2}}{\beta{\bar J}} e^{\beta{\bar J}z^2M^2/2}\\
&&\times\int_{0}^{\infty}dx\, x \left[I_0(x)\right]^{(\lambda/a)^2}I_0(z M x)e^{-x^2/(2\beta{\bar J})},
\ea
where $I_0(x)$ is the modified Bessel function of the first kind. Finally, the self-consistency condition
implies $M^2=(\beta{\bar J})^{-1}\partial\ln Z_{MF}/\partial z$, from which follows
\be
\label{M}
M\!=\frac{1}{(1+z)\beta{\bar J}}\frac
{\int_{0}^{\infty}\!dx x^2 \left[I_0(x)\right]^{(\lambda/a)^2}\!I_1(z M x)e^{-x^2/(2\beta{\bar J})}}
{\int_{0}^{\infty}\!dx x \left[I_0(x)\right]^{(\lambda/a)^2}\!I_0(z M x)e^{-x^2/(2\beta{\bar J})}}.
\ee

Near the mean-field transition temperature, $T_{c,MF}$, $M\rightarrow 0$ and we may approximate $I_0(z M x)\approx 1$
and $I_1(z M x)\approx z M x$. This, together with the parameterization $\beta_c{\bar J}=c(a/\lambda)^2$
for the inverse critical temperature, turns Eq. (\ref{M}) into
\be
\label{Tceq}
1=\frac{z}{1+z}\frac
{\int_{0}^{\infty}dx x^3 \left[I_0\left(\sqrt{2c}(a/\lambda) x\right)\right]^{(\lambda/a)^2}e^{-x^2}}
{\int_{0}^{\infty}dx x \left[I_0\left(\sqrt{2c}(a/\lambda) x\right)\right]^{(\lambda/a)^2}e^{-x^2}}.
\ee
Since $I_0(x\ll 1)\approx 1+x^2/4$ one finds in the limit $a/\lambda\ll 1$
that $\left[I_0\left(\sqrt{2c}(a/\lambda) x\right)\right]^{(\lambda/a)^2} \approx e^{cx^2/2}$. The
integrals in Eq. (\ref{Tceq}) are then readily evaluated with the result $c=2/(1+z)$, leading to
\be
\label{TcMF}
T_{c,MF}=\frac{1+z}{2}\left(\frac{\lambda}{a}\right)^2{\bar J}\sim J\ln(\lambda/a).
\ee

The phase stiffness of the coarse-grained model is determined by two types of processes: intra-block fluctuations
that reduce $M$ and with it the average coupling between superspins, and inter-block fluctuations of the superspins.
Our mean-field treatment ignores the second type of fluctuations. Fig. \ref{rhoT} depicts a fit to $\rho_s(T)/J$
of Hamiltonian (\ref{XYmodel}) using $M^2(T)$ obtained by solving Eq. (\ref{M}). The fit begins to deviate
from $\rho_s(T)$ as the latter becomes of order $(2/\pi)T$. Hence, the following physical picture emerges:
The phase stiffness is rapidly reduced from its large bare value by fluctuations on scales smaller
than $\lambda$. These include both longitudinal and transverse vortex excitations which reach,
according to our numerical findings, much higher densities below $T_c$ as compared to the case of
the standard $XY$ model. It is this renormalization that is responsible for the $T_c$ scaling with
$\ln(\lambda/a)$. Once $\rho_s$ approaches $(2/\pi)T$ vortex fluctuations on scales larger that
$\lambda$ become important and drive the system through a BKT transition.

Before proceeding to discuss the quantum fluctuations let us consider the approximation we made in neglecting the
crossover to a slower decay of the coupling, Eq. (\ref{f}), for $r<\xi$. Based on the insights gathered above we
can easily modify the mean-field treatment to include this crossover by defining the average coupling according to
\ba
\nonumber
{\bar J}&\approx& J\left(\frac{a}{\lambda}\right)^2\left[\int_a^\xi d^2r\, \frac{e^{-r/\lambda}}{\xi r}
+\int_\xi^\infty d^2r\, \frac{e^{-r/\lambda}}{r^2}\right] \\
&\approx& J\left(\frac{a}{\lambda}\right)^2\left[1+\ln\left(\frac{\lambda}{\xi}\right)\right].
\ea
Since $\lambda\gg\xi$ this introduces only a small correction to the mean-field transition temperature, Eq. (\ref{TcMF}).

\subsection{Quantum phase fluctuations}
\label{quantum}

For small $t_\perp$ the short time phase dynamics on a single site is governed by the first term in the action (\ref{Stheta}),
which implies
\be
\langle e^{-i[\theta_i(\tau)-\theta_i(\tau')]}\rangle=e^{-2\Delta_0|\tau-\tau'|}.
\ee
This means that a site phase is essentially constant over a time of the order of $\Delta_0^{-1}$ and allows us to coarse
grain space-time into "needles" of length $\Delta_0^{-1}$ in the imaginary time direction. The phases of the needles interact
according to the last term in Eq. (\ref{Stheta}), resulting in a coarse grained phase action
\be
S_\theta=-\sum_{\tau,\tau'}\sum_{i,j} g(r_{ij},\tau-\tau')\cos[\theta_i(\tau)-\theta_j(\tau')],
\ee
where $g(r,\tau)=\Delta_0^{-2}K(r,\tau)$, and $\theta_i(\tau)$ denotes the phase of the needle centered around time $\tau$ at
site $i$.

The fact that $g(r,\tau)$ is long ranged in the time direction allows us to apply a similar mean-field approach to the
one employed above in order to estimate $T_c$. Now, however, we include the effects of both quantum and thermal fluctuations.
For this purpose we divide space-time into rods of length $\beta$ in the time direction and spatial area $l_T^2$. Each
phase within a rod is taken to interact with all the other phases in its rod and in the $z$ neighboring rods with a
coupling strength which is the average of $g(r,\tau)$ over a rod
\ba
\nonumber
\label{barg}
{\bar g}&=&\frac{1}{N_{\rm rod}}\sum_{\tau=-\beta/2}^{\beta/2}\sum_{r<l_T}g(r,\tau) \\
&=&\frac{1}{N_{\rm rod}} \frac{2\pi t_\perp^4 N_F}{\Delta_0^3}\int_a^{l_T}dr\, r f(r).
\ea
Here $N_{\rm rod}=\beta\Delta_0(l_T/a)^2$ is the number of needles within a rod and we used the
fact that the averaging along the time direction leads to the couplings $f(r)$, Eq. (\ref{f}),  encountered previously in the
context of the thermal fluctuations. According to the mean-field analysis preceding Eq. (\ref{TcMF}) criticality occurs when
$N_{\rm rod}{\bar g}=2/(1+z)$. As a result one obtains
\be
\label{Tcquant}
T_c\sim{\rm min}(\Delta_0,\frac{v_F}{a})\exp\left[-\frac{2\Delta_0^3}{(1+z)t_\perp^4 N_F a^2}\right],
\ee
both in the clean and diffusive limits. We conclude that for $t_\perp\ll \sqrt{t\Delta_0}$, where our treatment applies,
quantum phase fluctuations induce an exponential suppression of $T_c$ from its value based on thermal fluctuations only,
Eq. (\ref{Tcclass}). Secondly, within our approximation the disorder has no effect on $T_c$ as long as the metallic
layer is in the diffusive regime.

\section{Large and intermediate $t_\perp$}

\subsection{The large $t_\perp$ limit}

When $t_\perp\gg U,t$ the physics is dominated by the inter-layer tunneling and the energy is
minimized by the creation  of a symmetric state on each $c-f$ dimer (assuming that the system is below
half filling). Denoting by $a_{i\sigma}$ the annihilation
operator of this state on site $i$ one finds
$c_{i\sigma}\approx f_{i\sigma}\approx a_{i\sigma}/\sqrt{2}$ and a disordered Hubbard model as the
effective Hamiltonian
\ba
\label{largeH}
\nonumber
&&\!\!\!\!\!\!\!\!\!\!\!\!\!\!\!H=-\frac{t}{2}\sum_{\langle i,j \rangle,\sigma}\left(a_{i\sigma}^{\dagger}
a_{j\sigma} + {\rm H.c}\right) \\
&&\hspace{-0.08cm}+\sum_{i\sigma}\left(\frac{\epsilon+V_i}{2}-\mu-t_\perp\right) n_{i\sigma}
-\frac{U}{4}\sum_i  n_{i\uparrow}n_{i\downarrow},
\ea
where $n_{i\sigma}=a_{i\sigma}^{\dagger}a_{i\sigma}$. In the weak coupling limit $U\ll t$ the phase
stiffness is largely determined by the amplitude of the order parameter. Hence, the BKT critical temperature
is very close to the BCS mean field transition temperature\cite{Halperin79} and Anderson's theorem applies.
For stronger interaction $U\sim t$ the model was investigated using both the mean-field
approximation\cite{Ghosal98} and quantum Monte-Carlo (QMC) simulations\cite{Scalettar99}. These studies
demonstrated how with increasing disorder strength the system becomes dominated by phase fluctuations
which eventually turn it into an insulator. In the limit $U\gg t$ the model
can be mapped\cite{DePalo99,Benfatto04} to a nearest-neighbor quantum $XY$ model whose $T_c\sim t^2/U$.

\subsection{The intermediate $t_\perp$ regime}

The preceding analysis shows that $T_c$ rises with small $t_\perp$ according to Eq. (\ref{Tcquant}),
while approaching an asymptotic large $t_\perp$ value which scales as $t\exp(-16t/U)$ and $t^2/U$ in the
weak and strong interacting limits, respectively. The absence of a small parameter in the intermediate regime
$t_\perp\sim U$ makes it a more difficult theoretical challenge and one needs to resort to numerical methods.
For this purpose we took advantage of the fact that the bilayer model is free from the sign
problem at all doping levels, and implemented a determinantal QMC technique to calculate
its phase stiffness. Consequently, $T_c$ was evaluated via the BKT criterion. The phase stiffness was extracted from
imaginary time current-current correlations according to a theorem by Scalapino, White and Zhang\cite{SWZ}
\be
\label{rhos}
\rho_s=-\frac{1}{4}\left[\langle K_x\rangle + \Lambda_{xx}(q_x=0,q_y\rightarrow 0,i\omega_n=0)\right].
\ee
Here
\be
K_x=-\frac{t}{N}\sum_{{\bf r},\sigma}\left( c_{{\bf r}+\hat{x},\sigma}^\dagger
c_{{\bf r},\sigma}+c_{{\bf r},\sigma}^\dagger c_{{\bf r}+\hat{x},\sigma}\right),
\ee
is half the kinetic energy per site, and in practice the limit $q_y\rightarrow 0$ of the correlation function
\be
\Lambda_{xx}({\bf q},i\omega_n)=\frac{1}{N}\int_0^\beta d\tau e^{i\omega_n\tau}\langle
j_x({\bf q},\tau)j_x({\bf -q},0)\rangle,
\ee
where
\be
j_x({\bf q})=it\sum_{{\bf r},\sigma}e^{-i{\bf q}\cdot{\bf r}}\left( c_{{\bf r}+\hat{x},\sigma}^\dagger
c_{{\bf r},\sigma}-c_{{\bf r},\sigma}^\dagger c_{{\bf r}+\hat{x},\sigma}\right),
\ee
stands for its value for $q_y=2\pi/L$ in the finite systems that we simulate. We used the BSS
algorithm\cite{BSS} to carry out the evolution in configuration space and the Hirsch
method\cite{Hirsch} for the measurement of $\Lambda_{xx}$. We found that in the range $U>t$
it was sufficient to set the QMC time slice to $\Delta\tau=(4U)^{-1}$ and sweep through the
system 5000-10,000 times in order to limit the systematic and statistical errors of the
calculated $T_c$ to a few percents.

\begin{figure}[t]
\includegraphics[angle=0,width=\linewidth,clip=true]{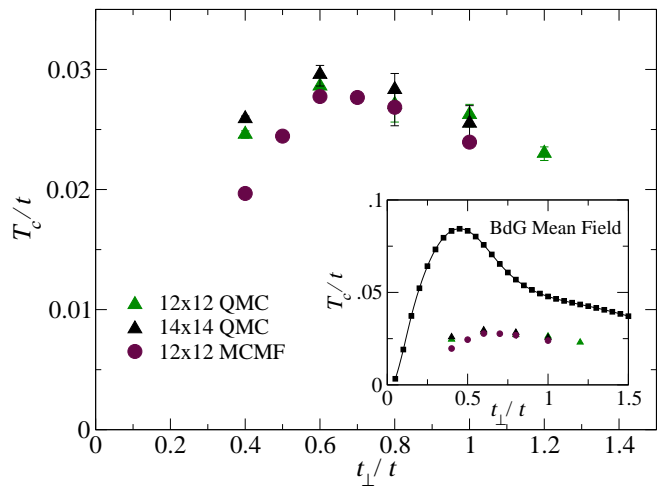}
\caption{The critical temperature as function of $t_\perp$ for a clean bilayer with $U/t=1$ and $n=1.5$,
calculated using QMC and MCMF methods. The inset depicts a comparison of these results with the
predications of the BdG mean-field theory.}
\label{Tcvstperp}
\end{figure}

Our results for a clean bilayer with $U/t=1$, total density $n=1.5$ and $L/a=12,14$ appear in
Fig. \ref{Tcvstperp}. A maximum in $T_c$ as function of $t_\perp$ is observed at $t_{\perp,{\rm max}}\simeq 0.6t$.
While this behavior is akin to the BdG mean-field findings of Ref. \onlinecite{Ours}, shown in the inset,
the two sets of data differ quantitatively. The peak in $T_c$, as calculated by QMC, occurs at somewhat higher
values of $t_\perp$ and its magnitude, $T_{c,{\rm max}}$, is about 3 times smaller than the corresponding
maximum in the BdG mean-field $T_c$.

\begin{figure}[t]
\includegraphics[angle=0,width=\linewidth,clip=true]{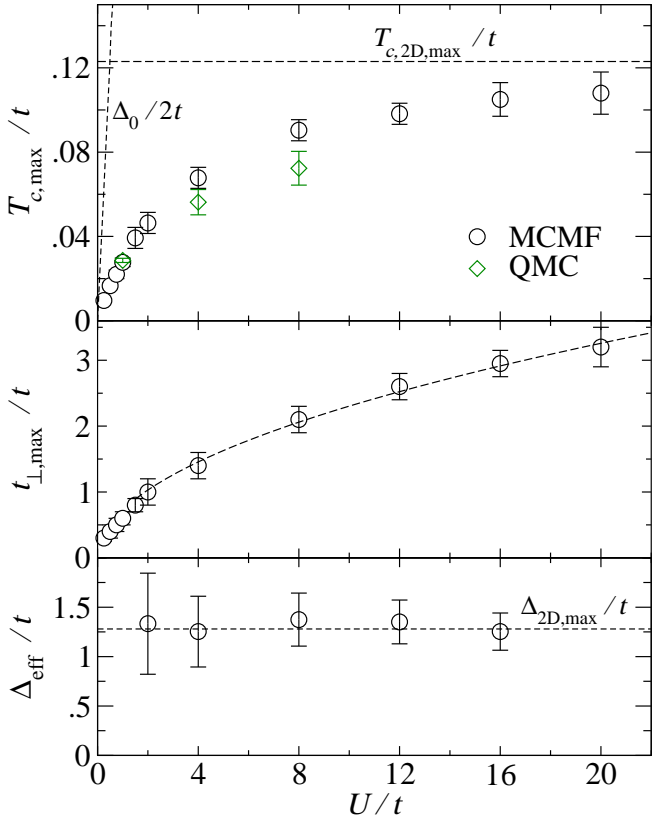}
\caption{Upper panel: The maximal critical temperature as function of $U/t$ for a clean $12\times 12$ bilayer with $n=1.5$.
The horizontal dashed line depicts the highest $T_c$ (maximized over $U/t$) of the single-layer attractive Hubbard model
with $n=0.5$. The left dashed line corresponds to the mean-field transition temperature of the disconnected Hubbard sites.
Middle panel: $t_{\perp,{\rm max}}$ for which $T_{c,{\rm max}}$ is obtained as function of $U/t$. The dashed line is a fit
to $t_{\perp,{\rm max}}\propto\sqrt{tU}$. Lower panel: $\Delta_{\rm eff}(t_{\perp,{\rm max}})=t_{\perp,{\rm max}}^2/\Delta_0$
as function of $U/t$. The dashed line corresponds to the average amplitude of the order parameter at the highest $T_c$ of
the single-layer attractive Hubbard model with $n=0.5$.}
\label{Uall}
\end{figure}

The use of QMC to study the $U/t$ dependence of $T_{c,{\rm max}}$ is restricted by the low temperatures that are
encountered in the small $U/t$ regime and the short $\Delta\tau$ required when $U/t$ is large. To partially overcome
these difficulties we employ an approximate method which neglects quantum fluctuations of the order parameter.
As we saw in Sec. \ref{thermal} such an approximation fails for small $t_\perp$. However, our findings demonstrate
that it is sufficient near the maximal $T_c$. In this Monte Carlo Mean Field method (MCMF), (originally introduced by
Mayr \etal\cite{Mayr}  in its $d$-wave version), a classical Monte-Carlo scheme is used to average over random
configurations of the local pairing field. The partition function is given by
\be
{\cal Z}=\int \prod_{i=1}^N d\Delta_i d\Delta_i^* e^{-(\beta/U)\sum_i|\Delta_i|^2}Z(\{\Delta_i\}),
\ee
where $Z(\{\Delta_i\})$ is the partition function of the BdG quasi-particles for a
given configuration of the pairing field. Observables, such as the phase stiffness, are calculated
using the quasi-particle Green's functions and averaged over the pairing field configurations.
The critical temperature is determined using Eq. (\ref{rhos}), just as in the QMC simulations.
As shown in Figs. \ref{Tcvstperp} and \ref{Uall} MCMF reproduces to within 20\% the QMC results
for $T_{c,{\rm max}}$ over the range $1\leq U/t \leq 8$ (we attribute the fact that the MCMF $T_c$
curve lies below the corresponding QMC curve in the $U/t=1$ system to a different finite size scaling
of the two methods). We therefore conclude that the maximal $T_c$ is largely governed by thermal
fluctuations.

Utilizing the MCMF method to calculate $T_{c,{\rm max}}$ beyond the $U/t$ range which is amenable to
QMC simulations one obtains the following behavior, depicted in Fig. \ref{Uall}. First, at small $U/t$
the maximal critical temperature scales with the mean-field temperature, $T_{\rm MF}\approx U/4$, of the
disconnected Hubbard sited. The latter is the temperature at which the pairing gap on each site closes and
thus sets a maximum conceivable value for $T_c$ which takes full advantage of the pairing scale. Numerically,
we find for $U<t$ that $T_{c,{\rm max}}\approx T_{\rm MF}/3$. This is to be contrasted with the mean-field
result of Ref. \onlinecite{Ours}, $T_{c,{\rm max}}\rightarrow T_{\rm MF}$ in the same range of parameters.
Presumably, the difference stems from the absence of (predominantly classical) phase fluctuations in the
mean-field calculation.

At large $U/t$, $T_{c,{\rm max}}$ seems to saturate to a limiting value, which is close to the maximal
$T_c$ of the quarter filled two-dimensional attractive Hubbard model.\cite{Tcmax-comm} 
Our MCMF results indicate that amplitude fluctuations of the order parameter play a minor role at the $T_c$
maximum of both models, changing it by about 5\%. We would therefore attempt to understand the above behavior
from the perspective of thermal phase fluctuations only.
To this end we decouple the interaction term in the original action (\ref{model}) and
integrate out the pairing sites while allowing only spatial phase fluctuation in the pairing field, \ie
assuming $\Delta_i(\tau)=\Delta_0e^{i\theta_i}$. The result is
\ba
\label{Sc}
\nonumber
S&=&-\beta\sum_{i,j,\sigma}\sum_{\omega_n}c_{i\sigma}^\dagger(\omega_n)\left[{\cal G}_{ij}^{-1}(\omega_n)
+\frac{t_\perp^2(i\omega_n-\mu)\delta_{ij}}{\omega_n^2+\mu^2+\Delta_0^2}\right]\\
\nonumber
&\times&c_{j\sigma}(\omega_n)+\beta\Delta_{\rm eff}\sum_i\sum_{\omega_n}\frac{\Delta_0^2}{\omega_n^2+\mu^2+\Delta_0^2} \\
&\times&\left[e^{i\theta_i}c_{i\uparrow}^\dagger(\omega_n)c_{i\downarrow}^\dagger(-\omega_n)+{\rm H.c.}\right],
\ea
where
\be
\label{deltaeff}
\Delta_{\rm eff}=\frac{t_\perp^2}{\Delta_0}.
\ee
As shown in Fig. \ref{Uall}, $t_{\perp,{\rm max}}$ scales as $\sqrt{tU}$ for $U>t$. Since $\Delta_0\approx U/2$
this implies that at $t_{\perp,{\rm max}}$ and for $U\gg t$ the ${\cal O}(t_\perp^2)$ correction to the first term in the action
(\ref{Sc}) can be neglected in comparison to ${\cal G}_{ij}^{-1}$. Moreover, since the important contribution to the
action comes from frequencies within the metallic bandwidth the prefactor inside the second term of Eq. (\ref{Sc}) can
be replaced in the limit $U\gg t$ by 1. Within these approximations Eq. (\ref{Sc}) becomes the action of a two-dimensional
attractive Hubbard whose filling is set by the filling of the metallic layer and whose pairing field amplitude is given by $\Delta_{\rm eff}$.
The lower panel of Fig. \ref{Uall} demonstrates that in the $U>t$ regime $\Delta_{\rm eff}(t_{\perp,{\rm max}})$ coincides with the
order parameter amplitude of the two-dimensional attractive Hubbard model at its maximum $T_c$ (as obtained using MCMF).
In other words, for large $U$ the bilayer achieves its optimal $T_c$ by adjusting the inter-layer
tunneling $t_\perp$ to a point that maps the bilayer onto a single layer attractive Hubbard with an optimal $U/t$ ratio.
Since the average pairing amplitude of the latter is of order $t$ Eq. (\ref{deltaeff}) implies the $t_{\perp,{\rm max}}\sim\sqrt{tU}$
scaling mentioned above.

Finally, in order to obtain a rough idea of the effect of disorder on $T_{c,{\rm max}}$ we resort to the mean-field
treatment of Ref. \onlinecite{Ours}. We apply it to a bilayer with $U=t$ where in the absence of disorder it differs
from the QMC and MCMF results by a factor of 3, see Fig. \ref{Tcvstperp}. The mean-field approximation consists of
decoupling the interaction term
$-Uf^\dagger_{i\uparrow}f_{i\uparrow}f^\dagger_{i\downarrow}f_{i\downarrow}
\rightarrow -\Delta^*_i f_{i\uparrow}f_{i\downarrow}-\Delta_i f^\dagger_{i\downarrow}f^\dagger_{i\uparrow}$,
and solving the BdG equations for the disordered bilayer at finite temperature. This is done under the self-consistent
condition $\Delta_i=U\langle f_{i\uparrow}f_{i\downarrow}\rangle$ and with a phase twist in the $x$ direction,
which enters the kinetic part of the Hamiltonian according to
\be
\label{subs}
-t\sum_{\langle i,j\rangle,\sigma}c^\dagger_{i\sigma}c_{j\sigma}\rightarrow
-t\sum_{\langle i,j\rangle,\sigma}e^{q_x(x_i-x_j)/2a}c^\dagger_{i\sigma}c_{j\sigma}.
\ee
The free energy, $F$, calculated from the BdG solutions, is then used to evaluate the bare phase
stiffness $\rho_s^0(T)=(a/L)^2\partial^2F/\partial q_x^2$, which includes the physics of thermally
excited quasiparticles but ignores the renormalization of the stiffness by phase fluctuations.

\begin{figure}[t]
\includegraphics[angle=0,width=\linewidth,clip=true]{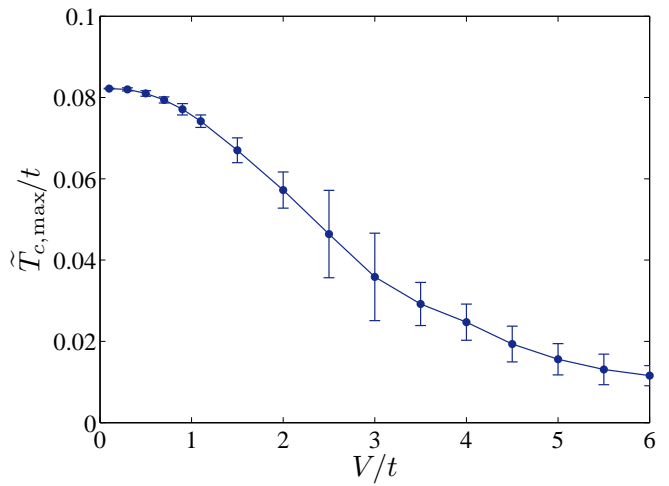}
\caption{The estimated maximal critical temperature ${\widetilde T}_{c,{\rm max}}$
deduced from the approximated BKT criterion ${\widetilde T}_{c,{\rm max}}=(\pi/2)\rho_s^0
({\widetilde T}_{c,{\rm max}})$ as function of the disorder strength $V$ for $U/t=1$
and $n=1.5$. The bars depict the standard deviation of the distribution of
${\widetilde T}_{c,{\rm max}}$. Based on our results for the clean system we expect the
actual maximal $T_c$ to be about 3 times smaller than the values presented here. For $V/t\gtrsim 2$
strong localization effects likely cause $T_c$ to drop even faster and to eventually vanish
at a critical disorder strength.}
\label{Tcmax}
\end{figure}

Figure \ref{Tcmax} depicts the estimated maximal critical temperature obtained from the
approximate BKT criterion ${\widetilde T}_{c,{\rm max}}=(\pi/2)\rho_s^0
({\widetilde T}_{c,{\rm max}},t_{\perp,{\rm max}})$ for a bilayer with $L=10a$, and $n=1.5$.
The results were averaged over up to 200 realizations of disorder in which every $V_i$
was drawn independently from a uniform distribution in the range $[-V,V]$.
As shown, the disorder has little effect on $T_{c,{\rm max}}$ as long as it is smaller than
the metallic hopping amplitude. For $V/t\gtrsim 2.5$ we found that the system's dimensionless
conductance, $g$, obeys $g<1$, indicating that strong localization effects become important
in this range. Such effects enhance phase fluctuations and are expected to induce a transition
to an insulating phase, which is not seen in the small system considered by us.
We also found that the value of $t_{\perp,{\rm max}}$ is not affected by the disorder
and remains fixed at $t_{\perp,{\rm max}}\approx0.48 t$.

\section{Conclusion}

Superconductivity contains within itself a built-in tension between its two essential prerequisites:
pairing and phase coherence. In all known examples where interactions are strong such that they lead to a
large pairing scale a concomitant suppression of the superfluid stiffness, and with it of $T_c$, takes place.
Canonical examples are the strong interaction regime of the attractive Hubbard model and the underdoped
cuprate superconductors. It was suggested that a possible way out of the dilemma is to separate the pairing
medium from the conduction electrons, thus maintaining their large phase stiffness.\cite{StevePhysica}
This strategy was pursued in Ref. \onlinecite{Ours} and its qualitative feasibility was demonstrated.
However, the consequences of superconducting phase fluctuations on the induced stiffness at the
interface between the two subsystems were not considered.

In the present work we have shown that phase fluctuations lead to an exponential suppression of the superfluid
stiffness when the interlayer coupling is small. For intermediate coupling these fluctuations cause a reduction
of $T_c$ by factor 3 relative to its value based solely on thermal excitation of quasiparticles. We believe
that our results have broader implications beyond the specific model studied here, as they point to the fact
that renormalization of the stiffness by phase fluctuations is important in systems with long range phase
couplings.

Despite the renormalization, long phase couplings are preferable to short ones from the perspective of
increasing $T_c$ at the surface of a phase fluctuating superconductor. As we showed, when a metal is overgrown
on the surface of a superconductor, additional phase couplings between sites on the interface are established
up to distances of the order of the metallic thermal length. This is true whether the added metal forms a two-dimensional
layer or is thicker than the thermal length and can therefore be considered as three dimensional. However,
within this range the decay of the coupling, technically described by the cooperon, follows $r^{-d}$ in the
$d$-dimensional case. This is a result of the fact that in three dimensions the electrons that mediate the
coupling spend part of their time moving in the direction perpendicular to the surface, thereby lowering
their probability to reach longer distances along the surface within the allotted coherence time $1/T$.
Consequently, our analysis suggests that one should attempt to make the metallic layer as thin as possible
in order to maximize the induced phase couplings and with them $T_c$.

Our results show that disorder has little effect on $T_c$, as long as it is weak enough not to cause strong
localization. From a practical point of view, a particular form of disorder, namely interface roughness, may
actually benefit the enhancement of $T_c$.\cite{Goren} This statement stems from the fact that when the pairing
scale is of the order of the metallic bandwidth maximal enhancement occurs at values of the interlayer
tunneling which are comparable to the metallic hopping amplitude. Such strong tunneling
across the interface may be difficult to achieve. An example can be found in the cuprate bilayers\cite{Yuli,Gozar}
where the intrinsic in-plane hopping is much stronger than the inter-plane one. Nevertheless, if
the interface is not perfect so that the pairing medium and the metal locally interpenetrate each other
electrons may tunnel between the two laterally, exploiting the large hopping amplitude in this direction.

\acknowledgments{We would like to thank T.~Paiva for useful comments.
This work was supported by the United States - Israel Binational Science Foundation (Grant No. 2008085)
and by the joint German-Israeli DIP project.}

\appendix

\section{Derivation of the phase action}

Here we provide details concerning the derivation of the phase action, Eq. (\ref{Stheta}). As discussed
in Section \ref{phase-action} we assume that the superconducting amplitude is fixed at its zero temperature
value $|\Delta_i(\tau)|=\Delta_0$ and proceed to write Eqs. (\ref{S1},\ref{MVreal}) in frequency space. In terms
of $\Psi_i^\dagger(\omega_m)=[f_{i\uparrow}^\dagger(\omega_m),f_{i\downarrow}(-\omega_m)]$ they become
\be
S=\sum_{i,j}\sum_{\omega_m,\omega_n}\Psi^\dagger_i(\omega_m)\left[M^{(0)}+V^{(0)}+V^{(1)}\right]
\Psi_j(\omega_n),
\ee
where
\ba
\label{MVk}
\nonumber
&&\hspace{-0.5cm}M^{(0)}_{ij}(\omega_m,\omega_n)=\beta[-i\omega_m{\bf I}-\mu\mbox{\boldmath $\sigma$}_3+\Delta_0\mbox{\boldmath $\sigma$}_1]\delta_{ij}\delta_{\omega_m,\omega_n},\\
\nonumber
&&\hspace{-0.5cm}V^{(0)}_{ij}(\omega_m,\omega_n)=\frac{\beta}{2}(\omega_m-\omega_n)\theta_i(\omega_m-\omega_n)\mbox{\boldmath $\sigma$}_3
\delta_{ij}\delta_{\omega_m,\omega_n},\\
\nonumber
&&\hspace{-0.5cm}V^{(1)}_{ij}(\omega_m,\omega_n)=\frac{\beta}{2} t_\perp^2 \sum_{\Omega_n}
\left[({\bf I}+\mbox{\boldmath $\sigma$}_3){\cal G}_{ij}(\omega_m-\Omega_n)F_i(\Omega_n)\right. \\
\nonumber
&&\hspace{-0.5cm}\times F_j^*(\Omega_n+\omega_n-\omega_m)-({\bf I}-\mbox{\boldmath $\sigma$}_3)
{\cal G}_{ji}(-\omega_n-\Omega_n)F_j(\Omega_n) \\
&&\hspace{-0.5cm}\left. \times F_i^*(\Omega_n+\omega_n-\omega_m)
\right].
\ea
Here, and throughout, $\omega_n$ and $\Omega_n$ are fermionic and bosonic Matsubara frequencies, respectively. We also denoted
\be
F_i(\tau)=e^{-i\theta_i(\tau)/2}=\sum_{\Omega_n}F_i(\Omega_n)e^{-i\Omega_n\tau}.
\ee

Our goal is to integrate out the fermions to obtain
\ba
\label{lntr}
\nonumber
S_{\theta}&=&-{\rm Tr}\ln[M^{(0)}+V^{(0)}+V^{(1)}] \\
\nonumber
&\approx&-{\rm Tr}\ln[M^{(0)}]-{\rm Tr}\left\{{M^{(0)}}^{-1}[V^{(0)}+V^{(1)}]\right\}\\
&&+\frac{1}{2}{\rm Tr}\left\{{M^{(0)}}^{-1}[V^{(0)}+V^{(1)}]\right\}^2.
\ea
$M^{(0)}$ is independent of $\theta$ and thus plays no role here. It is also trivial to see that the
linear term in $V^{(0)}$ vanishes. Hence, we turn our attention to
\ba
\label{first}
\nonumber
&&\hspace{-1cm}-{\rm Tr}\left[{M^{(0)}}^{-1}V^{(1)}\right] \\
&&\hspace{-1cm}\approx {\rm min}\left(1,\frac{8t}{\Delta_0}\right)\frac{t_\perp^2 N_F a^2}{T\Delta_0^2}
\sum_{i,\Omega_n}\Omega_n^2|\theta_i(\Omega_n)|^2,
\ea
evaluated for $\Omega_n<\Delta_0$. This, however, is negligible compared to
\ba
\label{timeder}
\nonumber
&&\hspace{-1cm}\frac{1}{2}{\rm Tr}\left[{M^{(0)}}^{-1}V^{(0)}{M^{(0)}}^{-1}V^{(0)}\right] \\
&&\hspace{-1cm}=\frac{1}{8T\Delta_0}\sum_{i,\Omega_n}\Omega_n^2|\theta_i(\Omega_n)|^2
+{\cal O}\left(\frac{\mu^2\Omega_n^2+\Omega_n^4}{T\Delta_0^3}\right),
\ea
which constitutes the first term in Eq. (\ref{Stheta}).

The term $\frac{1}{2}{\rm Tr}\left[{M^{(0)}}^{-1}V^{(1)}{M^{(0)}}^{-1}V^{(1)}\right]$ contains two contributions.
The first is
\ba
\label{C1}
\nonumber
&&\!\!\!\!\!C_1=-\sum_{i,j}\sum_{\Omega_m,\Omega_n}
\sum_{\omega_m,\omega_n}\frac{t_\perp^4\Delta_0^2}{(\omega_m^2+\mu^2+\Delta_0^2)(\omega_n^2+\mu^2+\Delta_0^2)} \\
\nonumber
&&\;\;\times \;F_i(\Omega_n)F_j^*(\Omega_n+\omega_m-\omega_n){\cal G}_{ij}(-\omega_m-\Omega_n)\\
&&\;\;\times \;F_i(\Omega_m)F_j^*(\Omega_m+\omega_n-\omega_m){\cal G}_{ij}(\omega_m-\Omega_m).
\ea
Its evaluation requires the disorder average $P_{ij}(\omega_n,\Omega_n)=\left\langle {\cal G}_{ij}\left(\omega_{n}\right)
{\cal G}_{ij}\left(\omega_{n}+\Omega_n\right)\right\rangle$. The metallic layer is assumed to contain Gaussian disorder
obeying $\langle V_i\rangle=0$ and $\langle V_i V_j\rangle =n{\cal V}^{2}\delta_{ij}$, where $n$ is the
average number of impurities per site and ${\cal V}^2=\langle V_i^2 \rangle/n$. The disorder scattering
is characterized by the elastic mean free time $\tau_{e}=[2\pi N_{F}a^{2}n{\cal V}^{2}]^{-1}$ and is
assumed weak, $k_Fl_e\gg 1$. The leading contribution to the above average comes from the cooperon,
\ie, the sum of ladder diagrams whose legs are constructed from disorder-averaged Green's functions
and whose rungs are disorder lines. In momentum space the cooperon takes the form\cite{Akkermans}
\ba
\label{P}
\nonumber
P({\bf Q},\omega_n,\Omega_n)&=&\frac{P_0({\bf Q},\omega_n,\Omega_n)}{1-(2\pi N_F a^2\tau_e)^{-1}P_0({\bf Q},\omega_n,\Omega_n)}.\\
\ea
Here
\ba
\label{P0int}
\nonumber
&&\hspace{-0.7cm}P_0({\bf Q},\omega_n,\Omega_n)=\frac{1}{N}\sum_{{\bf K}}\frac{1}
{-i\omega_{n}-\xi_{{\bf k}}-\frac{i}{2\tau_{e}}{\rm sign}(\omega_n)} \\
&&\hspace{-0.7cm}\times\frac{1}{i(\omega_n+\Omega_n)-\xi_{{\bf k}+{\bf Q}}+\frac{i}{2\tau_{e}}{\rm sign}(\omega_n+\Omega_n)},
\ea
with $\xi_{{\bf k}}=-2t[\cos(k_x a)+\cos(k_y a)]+\epsilon-\mu$, reduces to the cooperon of the clean system
in the limit $\tau_e\rightarrow\infty$.  For $Q=|{\bf Q}|\ll k_F$ it becomes
\be
\label{P0}
P_0({\bf Q},\omega_n,\Omega_n)=\pi N_Fa^2\tau_e\frac{\left|{\rm sign}(\omega_n+\Omega_n)-{\rm sign}(\omega_n)\right|}{\sqrt{(1+|\Omega_n|\tau_e)^2+(Ql_e)^2}}.
\ee
For a diffusive system, and as long as $Ql_{e}\ll1$ and $|\omega_n|\tau_{e}\ll1$,
one may expand the square root in Eq. (\ref{P0}), and plug the result into Eq. (\ref{P}) to find
\be
P({\bf Q},\omega_n,\Omega_n)=\pi N_Fa^2\frac{\left|{\rm sign}(\omega_n+\Omega_n)-{\rm sign}(\omega_n)\right|}{|\Omega_n|+DQ^2},
\ee
where $D=l_e^2/(2\tau_e)$ is the diffusion constant. At this point we would like to note that an additional time
scale, the phase breaking time $\tau_{\varphi}$, can be phenomenologically introduced into the problem and act
as a mass term for the cooperon. It is known\cite{AAK} that interactions in two dimensions lead to $\tau_{\varphi}^{-1}\sim(T\ln g)/g$,
where $g$ is the dimensionless conductance of the disordered layer. Our treatment of the disorder is valid for $g\gg 1$
and therefore $\tau_{\varphi}^{-1}\ll T$. Consequently, it does not alter our results and is not considered
further.

Transforming back to real space leads in the clean limit and $r_{ij}\gg k_F^{-1}$ to
\ba
\label{Pclean}
\nonumber
P_{ij}(\omega_n,\Omega_n)&=&\frac{1}{2}N_Fa^4\left|{\rm sign}(\omega_n+\Omega_n)-{\rm sign}(\omega_n)\right|\\
&\times&\frac{1}{v_F r_{ij}}e^{-|\Omega_m|r_{ij}/v_F},
\ea
while for the diffusive system and $r_{ij}\gg l_e$ the result is
\ba
\label{Pdiff}
\nonumber
P_{ij}(\omega_n,\Omega_n)&=&\frac{N_Fa^4}{2D}\left|{\rm sign}(\omega_n+\Omega_n)-{\rm sign}(\omega_n)\right|\\
&\times&K_0\left(r\sqrt{\frac{|\Omega_n|}{D}}\right),
\ea
where $K_0(x)$ is the Bessel function of the first kind.

Plugging Eq. (\ref{Pclean}) into Eq. (\ref{C1}), defining $\omega=\Omega_m-\omega_m$, $\omega'=\Omega_n+\omega_m$ and
carrying out the summation over $\omega_m$ and $\omega_n$ using the fact that at the relevant low temperature range
they can be approximated by integrals one obtains $C_1=\sum_{i,j}C_{1,ij}$, where in the clean limit
\ba
\label{C1ij}
\nonumber
&&\hspace{-0.42cm}C_{1,ij}=-\frac{t_\perp^4 N_F a^4T^2}{8v_F r_{ij}}\int_0^\beta\!\prod_{n=1}^4 d\tau_n
e^{-\Delta_0|\tau_1-\tau_2|} e^{-\Delta_0|\tau_3-\tau_4|} \\
\nonumber
&&\,\times\; F_i(\tau_1)F_i(\tau_2)F_j^*(\tau_3)F_j^*(\tau_4)\sum_{\omega,\omega'}\left|{\rm sign}(\omega')-{\rm sign}(\omega)\right|\\
&&\,\times\; e^{i[\omega(\tau_1-\tau_3)+\omega'(\tau_2-\tau_4)]-|\omega'-\omega|r_{ij}/v_F}.
\ea
For $r_{ij}>l_T=v_F/T$ the sum is dominated by the lowest frequency difference and $C_{1,ij}$ decays
as $(1/r_{ij})e^{-2\pi r_{ij}/l_T}$. For $r_{ij}\ll l_T$ the sums may be approximated by integrals with the result
\ba
\label{C1ij2}
\nonumber
&&\hspace{-0.42cm}C_{1,ij}=-\frac{t_\perp^4 N_F a^4}{8\pi^2 v_F r_{ij}}\int_0^\beta\!\prod_{n=1}^4 d\tau_n
e^{-\Delta_0|\tau_1-\tau_2|} e^{-\Delta_0|\tau_3-\tau_4|} \\
\nonumber
&&\hspace{0.4cm} \times\; \frac{(\tau_1-\tau_3)(\tau_2-\tau_4)+(\frac{r_{ij}}{v_F})^2}{[(\tau_1-\tau_3)^2+(\frac{r_{ij}}{v_F})^2]
[(\tau_2-\tau_4)^2+(\frac{r_{ij}}{v_F})^2]} \\
&&\hspace{0.4 cm}\times\; F_i(\tau_1)F_i(\tau_2)F_j^*(\tau_3)F_j^*(\tau_4).
\ea

Owing to Eq. (\ref{timeder}) $\langle F_i(\tau)F_i^*(\tau')\rangle=e^{-\Delta_0|\tau-\tau'|/2}$,
which means that $F_i$ may be considered constant within a time window of order $\Delta_0^{-1}$.
This fact and the exponential factors in Eq. (\ref{C1ij2}) enable us to approximate $F_i(\tau_1)=F_i(\tau_2)$
and $F_j^*(\tau_3)=F_j^*(\tau_4)$, leading to the coupling $\cos[\theta_i(\tau)-\theta_j(\tau')]$ mediated by
pair tunneling between the two sites, and constituting the second term in
the action, Eq. (\ref{Stheta}). The associated kernel may be evaluated in the range $a\ll r_{ij}\ll\xi=v_F/\Delta_0$
by taking $r_{ij}\rightarrow 0$ in Eq. (\ref{C1ij2}) and carrying out the integration over $\tau_2$ and $\tau_3$.
The result
\ba
\nonumber
&&\hspace{-0.4cm}K(r,\tau)=\frac{t_\perp^4 N_F a^4}{8\pi^2v_Fr}
\left[\left(e^{\Delta_0\tau}{\rm Ei}[-\Delta_0\tau]-e^{-\Delta_0\tau}{\rm Ei}[\Delta_0\tau]\right)^2 \right.\\
&&\hspace{0.79cm}+\left.\pi^2e^{-2\Delta_0|\tau|}\right],
\ea
where ${\rm Ei}(x)$ is the exponential integral function, is approximated by Eq. (\ref{Kshortclean}).
For $l_T \gg r_{ij}\gg \xi$ we may expand the integrand in Eq. (\ref{C1ij2}) to second order in $\tau_1-\tau_2$
and $\tau_3-\tau_4$. Integration over $\tau_2$ and $\tau_3$ then yields Eq. (\ref{Klongclean}).

Repeating the derivation using Eq. (\ref{Pdiff}) for the diffusive case produces a similar coupling with the
kernel
\ba
\nonumber
\label{Kdiffapp}
K(r,\tau)&=&\frac{t_\perp^4\Delta_0 N_F a^4 T}{\pi D}\sum_{\omega>0}
\frac{1}{\omega(\omega^2+\Delta_0^2)}K_0\left(r\sqrt{\frac{2\omega}{D}}\right) \\
\nonumber
&\times&\int_{-\infty}^\infty d\eta e^{-2\Delta_0|\eta|}
\frac{\sin[2\omega(\tau+\eta)]}{\tau+\eta} \\
&\times&\left[\omega\cos(2\omega\eta)+\Delta_0\sin(2\omega|\eta|)\right].
\ea
The exponential decay of $K_0(x)$ for $x>1$ induces the decay of the kernel when $r_{ij}>l_T=\sqrt{D/T}$.
This also means that for $l_T \gg r_{ij}\gg \xi=\sqrt{D/\Delta_0}$ only $\omega<\Delta_0(\xi/r)^2$ contribute.
Since the exponential factor in Eq. (\ref{Kdiffapp}) implies $|\eta|<\Delta_0$ we may expand the integrand in
$\omega|\eta|<1$  and evaluate the integrals with the approximated result Eq. (\ref{Klongdiff}).

The second contribution to the effective action coming from
$\frac{1}{2}{\rm Tr}\left[{M^{(0)}}^{-1}V^{(1)}{M^{(0)}}^{-1}V^{(1)}\right]$ is
\ba
\label{C2}
\nonumber
&&\!\!\!\!\!\!\!\!\!\!\!\!C_2=\sum_{i,j}\sum_{\Omega_m,\Omega_n}
\sum_{\omega_m,\omega_n}\frac{t_\perp^4(i\omega_m+\mu)(i\omega_n+\mu)}{(\omega_m^2+\mu^2+\Delta_0^2)(\omega_n^2+\mu^2+\Delta_0^2)} \\
\nonumber
&&\!\!\!\times \;F_i(\Omega_n)F_j^*(\Omega_n+\omega_m-\omega_n){\cal G}_{ij}(-\omega_m-\Omega_n)\\
&&\!\!\!\times \;F_j(\Omega_m)F_i^*(\Omega_m+\omega_n-\omega_m){\cal G}_{ji}(-\omega_n-\Omega_m).
\ea
Following a similar line of derivation to the one taken above we find that this contribution corresponds to a
coupling $\cos\left\{\frac{1}{2}[\theta_i(\tau)-\theta_j(\tau)-\theta_i(\tau')+\theta_j(\tau')]\right\}$,
induced by single-particle tunneling between the sites.\cite{schonreview} However, we find that it has a
negligible effect in comparison to the pair-tunneling term. For example in the clean system and for
$l_T \gg r_{ij}\gg \xi$ the associated kernel reads
\be
\label{tildeK}
{\tilde K}(r,\tau)=\frac{t_\perp^4 N_F a^4 \mu^2}{2\pi^2\Delta_0^4 v_F}\frac{1}{r}\frac{\tau^2-(r/v_F)^2}{[\tau^2+(r/v_F)^2]^2}.
\ee
Applying the mean field treatment of Sec. \ref{quantum} to this term yields an average coupling constant
${\bar g}\simeq(2t_\perp^4 N_F a^2\mu^2)/(N_{\rm rod}\pi\Delta_0^5)$ to be compared with
${\bar g}=(t_\perp^4 N_F a^2/N_{\rm rod}\Delta_0^3)\ln(l_T/\xi)$, Eq. (\ref{barg}).

Finally, the ${\rm Tr}\left[{M^{(0)}}^{-1}V^{(0)}{M^{(0)}}^{-1}V^{(1)}\right]$ piece in Eq. (\ref{lntr})
can be shown to result in a $(\partial_\tau\theta_i)^2$ term, which is of order $t_\perp^2$ and is therefore
insignificant compared to Eq. (\ref{timeder}).

\end{document}